\documentclass[usegraphicx]{aastex631}
\usepackage{amssymb}
\usepackage{bm}
\usepackage{color}
\usepackage{graphicx}

\input epsf

\def\Lya{Ly$\alpha$}
\def\Lyb{Ly$\beta$}

\def\HI{\hbox{H~$\scriptstyle\rm I\ $}}

\def\ltsima{$\; \buildrel < \over \sim \;$}
\def\lsim{\lower.5ex\hbox{\ltsima}}
\def\gtsima{$\; \buildrel > \over \sim \;$}
\def\gsim{\lower.5ex\hbox{\gtsima}}

\def\aap{A\&Ap}

\def\apjl{ApJ}
\def\apjs{ApJS}

\def\spose#1{\hbox to 0pt{#1\hss}}
\def\lta{\mathrel{\spose{\lower 3pt\hbox{$\mathchar"218$}}
     \raise 2.0pt\hbox{$\mathchar"13C$}}}
\def\gta{\mathrel{\spose{\lower 3pt\hbox{$\mathchar"218$}}
     \raise 2.0pt\hbox{$\mathchar"13E$}}}

\begin{document}

\title{Observational constraints on the metagalactic \Lya\ photon
  scattering rate at high redshift}

\author{Avery Meiksin}
\email{E-mail:\ meiksin@ed.ac.uk}
\affiliation{Institute for Astronomy, University of Edinburgh\\
  Royal Observatory of Edinburgh\\
  Edinburgh\ EH9\ 3HJ, UK}
\altaffiliation[Affiliate of\ ]{Scottish Universities Physics Alliance (SUPA)}

\date{\today}

\begin{abstract}
  The scattering of \Lya\ photons from the first radiating sources in
  the Universe plays a pivotal role in 21-cm radio detections of
  Cosmic Dawn and the Epoch of Reionization through the
  Wouthuysen-Field effect. New data from \emph{JWST} show the
  \Lya\ photon scattering rate exceeds that required to decouple the
  intergalactic hydrogen spin temperature from that of the Cosmic Microwave
  Background up to $z\sim14$ and render the neutral hydrogen visible.
\end{abstract}

\keywords{cosmology --  reionization -- intergalactic medium
}

\section{Introduction}
\label{sec:intro}

The reionization of intergalactic \HI\ is the last major phase change
in the baryonic component of the Universe. Two avenues have been
followed for its discovery:\ the search for the reionization sources
and the direct detection of the Epoch of Reionization (EoR) through
radio 21-cm measurements. The two are intimately related through the
production of both ionizing radiation and \Lya\ photons. The latter
are crucial to unpin the \HI\ spin temperature from the Cosmic
Microwave Background (CMB) through the Wouthuysen-Field effect (WFE)
\citep{1952AJ.....57R..31W,1959ApJ...129..536F}, and so render
detectable the EoR, and the Cosmic Dawn of the first radiating sources
leading up to it, in the radio against the CMB. While theoretical
predictions suggest galaxies provide sufficient \Lya\ photons for the
WFE to be effective at redshifts $z<20$, and possibly to $z<30$,
direct observational support for the required galaxies has been
constrained to $z<10$ \citep{2018MNRAS.480L..43M}. It is shown here
that recent deeper \emph{JWST} observations suggest galaxies provide
sufficient numbers of \Lya\ photons for the WFE to act at least to
$z\sim14$, the $3\sigma$ upper limit for the EoR from the
\emph{Planck} 2018 data.

\section{The WFE and the EoR}
\label{sec:wfeeor}
 
 The condition for the WFE to be effective against the CMB is
\begin{equation}
\frac{P_\alpha}{P_\mathrm{th}} = \frac{1}{18\pi}\frac{f_\mathrm{\alpha L}f_\mathrm{LH}}{f_\mathrm{esc}}n_\mathrm{H}\lambda_\alpha^3\frac{A_\alpha}{A_{10}}\frac{T_*}{T_\mathrm{CMB}}>1,
\label{eq:PaPth}
\end{equation}
\citep{1997ApJ...475..429M} where $\lambda_\alpha$ is the \Lya\ photon
wavelength, $A_\alpha$ and $A_{10}$ are the spontaneous decay rates of
the \Lya\ and 21-cm hyperfine transitions, respectively,
$T_*=h_\mathrm{P}\nu_{10}/k_\mathrm{B}$ where $\nu_{10}$ is the 21-cm
transition frequency, $h_\mathrm{P}$ is Planck's constant,
$k_\mathrm{B}$ is the Boltzmann constant, $T_\mathrm{CMB}$ is the
CMB temperature, and
$P_\mathrm{th}=(27/4)A_{10}T_\mathrm{CMB}/T_*$ is the thermalization rate. The cosmic number
densities of \Lya, $n_\alpha$, and Lyman Limit, $n_L$, photons
generated by galaxies are related through
$n_\alpha = f_\mathrm{\alpha L}n_L$ with
$f_\mathrm{\alpha L}\sim 1$. Only a fraction up to
$f_\mathrm{esc}\sim0.2$ of Lyman Limit photons escape into the
Intergalactic Medium (IGM) \citep{2022ARA&A..60..121R}. Here,
$f_\mathrm{LH}=f_\mathrm{esc}n_L/n_\mathrm{H}\sim0.01-1$
corresponds to the EoR, so that the combination
$(f_\mathrm{\alpha L}f_\mathrm{LH}/f_\mathrm{esc})>0.05$ during  
the EoR. For a baryon density $\Omega_bh^2=0.022$ and
$T_\mathrm{CMB}=2.725$~K today, during the EoR
$P_\alpha/P_\mathrm{th}>0.002(1+z)^2$ exceeds unity by $z\sim25$,  
corresponding to the 21-cm line redshifted to $\sim50$~MHz, making it
possible to detect the EoR in the low-frequency radio band
\citep{1997ApJ...475..429M}, and motivating radio EoR experiments
\citep{2012arXiv1212.3497E}.

\section{The metagalactic \Lya\ photon scattering rate}
\label{sec:MGLyA}

\begin{figure}
\scalebox{0.95}{\includegraphics{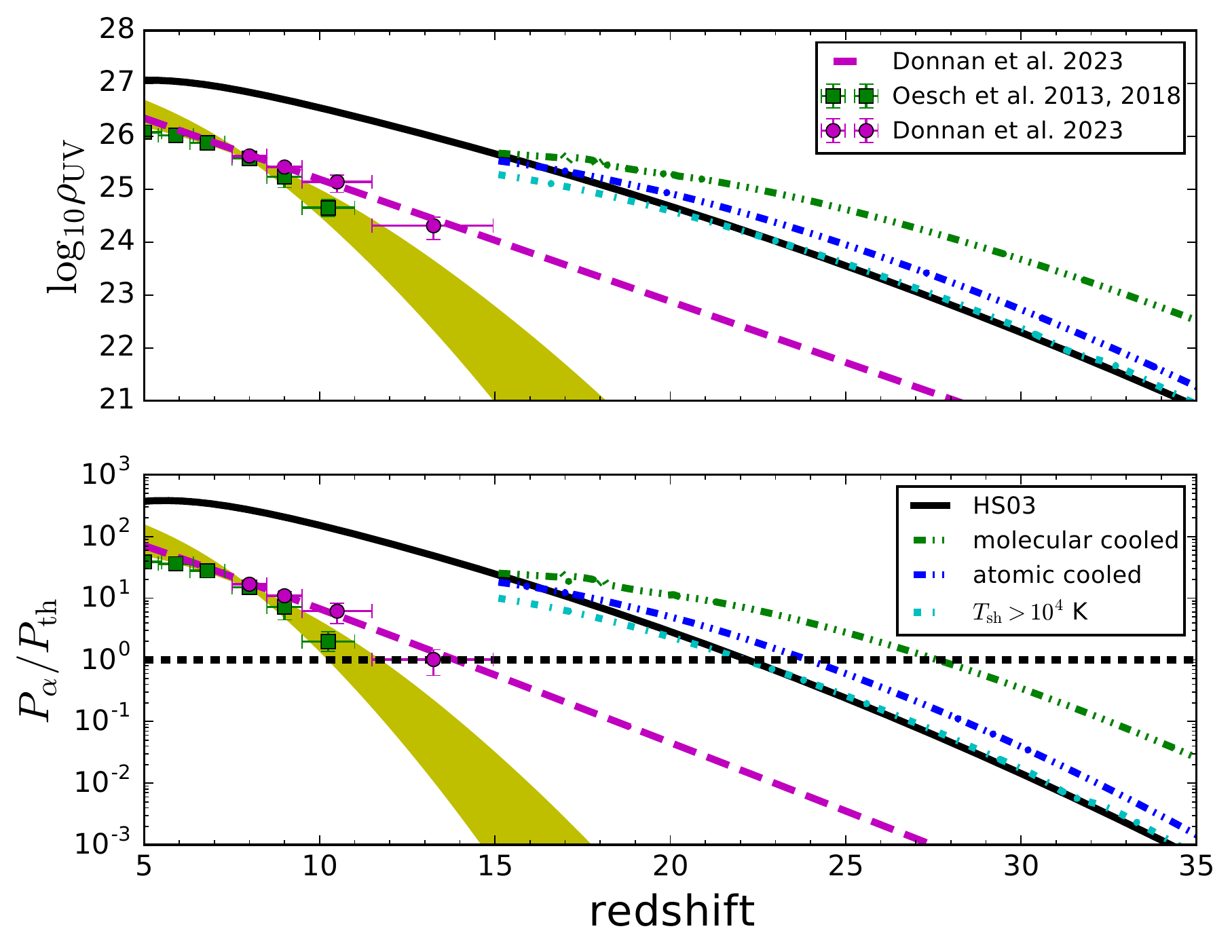}}
\caption{Evolution of the galactic UV luminosity density at 1500A ($\mathrm{ergs\,
  s^{-1}\, Hz^{-1}\, Mpc^{-3}}$)
  (upper panel) and the \Lya\ scattering rate $P_\alpha$, normalized
  by the thermalization rate $P_{\rm th}$ (lower panel).  The curves
  correspond to the indicated minimum halo mass thresholds required for
  star formation.
}
\label{fig:rhoUVPaPth_evol}
\end{figure}

The UV continuum radiation emitted by a galaxy between the \Lya\ and
\Lyb\ frequencies will be redshifted to the local \Lya\ frequency and
contribute to the WFE. In terms of the UV luminosity density of galaxies
$\rho_\mathrm{UV}$, the \Lya\ photon scattering rate is \citep[for a flat spectrum,
][]{2023MNRAS.518.6011D} $P_\alpha \simeq (5/27)\tau_\alpha\rho_\mathrm{UV}/(h_\mathrm{P}n_\mathrm{H})$,
where $\tau_\alpha$ is the Gunn-Peterson optical depth, which is also
the number of times a \Lya\ photon scatters before redshifting away
\citep{1959ApJ...129..536F, 2012MNRAS.426.2380H}.

For a Salpeter IMF and expected young galaxy metallicity, the cosmic
star formation rate density $\dot\rho_*\simeq K_\mathrm{UV}\rho_{UV}$,
where
$K_\mathrm{UV}\simeq1.15\times10^{-28}\,\mathrm{M_\odot\,
  yr^{-1}/(erg\, s^{-1}\, Hz^{-1})}$ \citep{2014ARA&A..52..415M}. A
simple estimate for $\dot\rho_*$ is given by the fraction
$F_\mathrm{gal}$ of haloes that collapse with masses above the
threshold required for star formation
\citep[eg][]{2005ApJ...626....1B}:\
$\dot\rho_* = \bar \rho_b \epsilon_*dF_\mathrm{gal}/dt$, where
$\bar \rho_b$ is the mean cosmic baryon density and $\epsilon_*$ is the
star formation efficiency. The thresholds for star-forming haloes are
taken as
$M_\mathrm{thresh}\simeq10^6[26/(1+z)]^{1/2}\,\mathrm{M_\odot}$ and
$M_\mathrm{thresh}\simeq9.1\times10^6\exp[-(1+z)/51]\,\mathrm{M_\odot}$
for molecular hydrogen and atomic hydrogen cooled haloes, respectively
\citep{2011MNRAS.417.1480M}. \citep[The latter applies if molecular
hydrogen formation is disrupted by the radiation from an earlier generation of galaxies, ][]{1997ApJ...476..458H}. A common
proxy for atomic-cooled haloes is to require their post-shock or viral
temperature to exceed $10^4$~K. The resulting UV luminosity densities
are shown in the upper panel of Fig.~\ref{fig:rhoUVPaPth_evol} for $\epsilon_*=0.01$,
adopting the halo mass function from \citet{2007MNRAS.374....2R},
adapted to \emph{Planck} 2018 constraints on the cosmological parameters
\citep{2018arXiv180706209P}. The estimate
compares well with a more sophisticated model, allowing star-formation
only in haloes with virial temperatures above $10^4$~K
\citep[][HS03]{2003MNRAS.341.1253H}, updated to the \emph{Planck} 2018
power spectrum normalization.

These are compared with the measured values from
\citet{2013ApJ...773...75O, 2018ApJ...855..105O} and
\citet{2023MNRAS.518.6011D} (for $M_{1500}<-17$), in the upper panel
of Fig.~\ref{fig:rhoUVPaPth_evol}. The inferred values for
$P_\alpha/P_\mathrm{th}$ are shown in the lower panel. By $z<10$, the
measured UV emissivity shows $P_\alpha>P_\mathrm{th}$, so that the
hydrogen spin temperature should be well removed from the CMB
temperature. The data from \citet{2013ApJ...773...75O,
  2018ApJ...855..105O}, however, suggest a rapidly declining
emissivity at $z>9$. The shaded region represents the declining number
of collapsed haloes with masses $9.5<\log_{10}M_h/M_\odot<10.5$. The
trend suggests by $z=13$, the data no longer ensure
$P_\alpha>P_\mathrm{th}$. The observations of
\citet{2023MNRAS.518.6011D} using \emph{JWST} show on the contrary,
$P_\alpha/P_\mathrm{th}>1$ is maintained to $z\sim14$. This is
sufficient to cover the entire waveband (115--203~MHz) probed by the
Low Frequency Array (LOFAR) High-band Antenna EoR experiment
\citep{2013A&A...556A...2V}.

\bibliographystyle{aasjournal}
\bibliography{ms}


\end{document}